\documentclass[preprint]{ptephy_v1}

\usepackage{hyperref}
\usepackage{amsmath}
\usepackage{comment}
\usepackage{ulem}

\newcommand{\Ba}{\mathrm{^{133}Ba}}
\newcommand{\Rn}{\mathrm{^{222}Rn}}
\newcommand{\Ra}{\mathrm{^{226}Ra}}
\newcommand{\gdso}{\mathrm{{Gd_{2}(SO_{4})_{3}\cdot 8H_{2}O}}}

\begin{document}

\title{Rapid Analysis of $\Ra$ in\\ Ultrapure Gadolinium Sulfate Octahydrate}

\author{Y.~Sakakieda}
\affil{University of Tsukuba, Graduate School of Science and Technology, 1-1-1 Tennodai, Tsukuba, Ibaraki, 305-8577, Japan}
\author[2,*]{K.~Hosokawa}
\affil{Kamioka Observatory, Institute for Cosmic Ray Research, University of Tokyo, Kamioka, Gifu 506-1205, Japan \email{hosokawa@km.icrr.u-tokyo.ac.jp}}
\author{F.~Nakanishi}
\affil{Okayama University, 3-1-1 Tsushima-naka, Kita-ku, Okayama, 700-8530, Japan}
\author[3]{Y.~Hino}
\author[2]{Y.~Inome}
\author{A.~Sakaguchi}
\affil{University of Tsukuba, Institute of Pure and Applied Sciences, 1-1-1 Tennodai, Tsukuba, Ibaraki, 305-8577, Japan}
\author[4]{Y.~Takaku}
\author[2]{M.~Ikeda}
\author[2]{H.~Sekiya}

\begin{abstract}
Numerous particle physics experiments utilize gadolinium (Gd), a rare earth element with the most significant neutron capture cross-section among all elements, to detect anti-neutrinos via inverse beta-decays or to remove neutron-induced background events.
For example, to load Gd into water Cherenkov detectors, $\gdso$ is dissolved and rare event search experiments are required to screen for radioactive impurities in $\gdso$ before dissolution.
This study developed a new method to rapidly measure the radium-226($\Ra$) concentration in $\gdso$.
This method requires only three days to measure a batch of samples, as opposed to the usual method using high-purity germanium detectors, which takes approximately 20 days after arrival.
The detection limit for the measurement of $\Ra$ is 0.32 mBq/kg.
This method has been already used for $\gdso$ screening at the Super-Kamiokande Gd(SK-Gd) project, and it can be applied to future experiments.
\end{abstract}
\subjectindex{H20}

\maketitle

\section {Introduction}
Gadolinium (Gd) is a rare earth element with the most significant thermal neutron capture cross-section among all elements. Upon neutron capture on Gd, gamma-rays are emitted with a total energy of approximately 8 MeV. 
As a result, Gd has various applications in particle physics experiments. For example, neutrino measurements, such as those performed by Daya Bay, Double-Chooz, JSNS$^2$, and RENO, utilize Gd-loaded liquid scintillators to improve their neutrino detection and background rejection. Tagging a subsequent neutron capture signal of an inverse beta decay reaction (IBD)~\cite{DayaBay, DoubleChooz, JSNS2, RENO} achieves this. Additionally, XENONnT plans to reduce the number of neutron-induced background events by detecting neutrons using Gd-loaded water surrounding the time projection chamber~\cite{XENON-nT}.
In a new experimental phase, Super-Kamiokande(SK)has loaded Gd into the ultrapure water, called SK-Gd~\cite{T1LoadingPaper}.
Improving IBD identification through this method is expected to enhance the search sensitivity for the diffused supernovae neutrino background.

SK-Gd aims to enhance the neutron detection efficiency by adding Gd to the detector, using $\gdso$ as a reliable source with minimal impact on detector materials.
The Gd-loading was conducted by dissolving approximately 13 tons of $\gdso$ between July and August 2020, followed by 26 tons between June and July 2022.
Since radioactive impurities can be sources of serious background events for low-energy neutrino studies, such as solar neutrinos, all production lots of $\gdso$ are screened for radioactive contaminants before dissolving them.
The screening mainly focused on uranium(U), thorium (Th), and actinium (Ac) decay series.
Chemical separation using a resin and inductively coupled plasma mass spectrometry (ICP-MS) is used to determine the levels of $\rm{^{238}U}$ and $\rm{^{232}Th}$ in the $\gdso$ used for SK-Gd. At the same time, high-purity germanium (HPGe) detectors in underground laboratories are employed to measure other radioactive impurities, including $\Ra$. The allowable level of $\Ra$ in the $\gdso$ used for SK-Gd is $<$0.5 mBq/kg, as specified in the study by~\cite{T1LoadingPaper}, to maintain the sensitivity for the $^8$B solar neutrino measurement. 
Evaluating the radioactive contamination in one $\gdso$ sample from a production lot using an HPGe detector takes approximately 20 days, making it time-consuming to assess all the lots. Additionally, 10 of the 20 days were spent waiting for a radioactive equilibrium of $\Rn$ ($\tau{1/2}=3.8$ days), which is a progeny nucleus of $\Ra$ and further slowed down the process. Therefore, a new evaluation method was developed to speed up the $\Ra$ concentration measurement.

The study is structured as follows: Section 2 introduces the materials and equipment for this study.
Section 3 describes the experimental method and its performance.
Section 4 presents an application of this method to the last two lots that were delivered only one week before their actual introduction into SK-Gd.
Finally, Section 5 concludes the paper.

\section{Materials and Equipment}
This study introduces a new method for measuring the ultra-trace concentration of $\Ra$ in an ultrapure $\gdso$.
To avoid contamination from impurities, we prepared high-quality reagents and thoroughly washed all tools used in the experiment, as described below.

\subsection{Chemical materials and equipment}
\label{sec:ReagentsAndEquip}

Ultra-high-purity $\gdso$ (Nippon Yttrium Co., Ltd) developed in \cite{T1ScreeningPaper} is used.

High-quality reagents and ultrapure water with a specific electrical resistivity of 18.2~M$\Omega\cdot$cm) are necessary for diluting solutions in this method.
Ultrapure-grade HNO$_3$(TAMAPURE-AA-100, Tama Chemicals Co., Ltd.), particular grade ethylenediaminetetraacetic acid diammonium salt(EDTA$\cdot \rm{2NH_4}$, DOJINDO LABORATORIES), and electronic grade NH$_4$OH(KANTO CHEMICAL CO., INC.) are selected.
To wash containers, columns, peristaltic tubes, and quartz beakers using electronic quality HNO$_3$, electronic-grade HCl(KANTO CHEMICAL CO., INC.), ultrapure-grade HF(TAMAPURE-AA-100, Tama Chemicals Co., Ltd.), or a combination thereof are used.

AnaLig$^\text{\textregistered}$ Ra-01 (IBC Advanced Technologies) resin.
This resin is known for its high absorption property for Ra ions, even from high-matrix samples such as high-concentration $\gdso$ solutions. The absorbed Ra isotopes are extracted with EDTA under basic conditions, as described in \cite{SakakiedaProc}.

Ln resin (LN-B25-A, Eichrom Technologies, LLC) uses di-ethylhexyl phosphoric acid as an extractant and can selectively extract various lanthanoid elements. The extraction selectivity can be altered by adjusting the nitric acid concentration~\cite{LnResin}.
This study employed the Ln resin for Gd removal, as described in Section~\ref{sec:Ln}.

A wide-mouth bottle with a capacity of 500 mL (AS ONE Corporation) and flat-base self-standing tubes with a volume of 5 mL (SARSTEDT) were used as solution containers. Quartz beakers were employed to contain the eluent due to their resistance to the heat generated during the pyrolysis and evaporation steps. M-size and S-size Muromac$^\text{\textregistered}$ mini-columns (MUROMACHI CHEMICALS INC.) were used to pack the AnaLig$^\text{\textregistered}$ and Ln resins, respectively.

During the $\Ra$ extraction step, a peristaltic pump MINIPULS$^\text{\textregistered}$ 3  (Gilson) was used to transfer the sample solution to the columns.

\begin{figure}[!t]
\centering\includegraphics[width=0.9\linewidth] {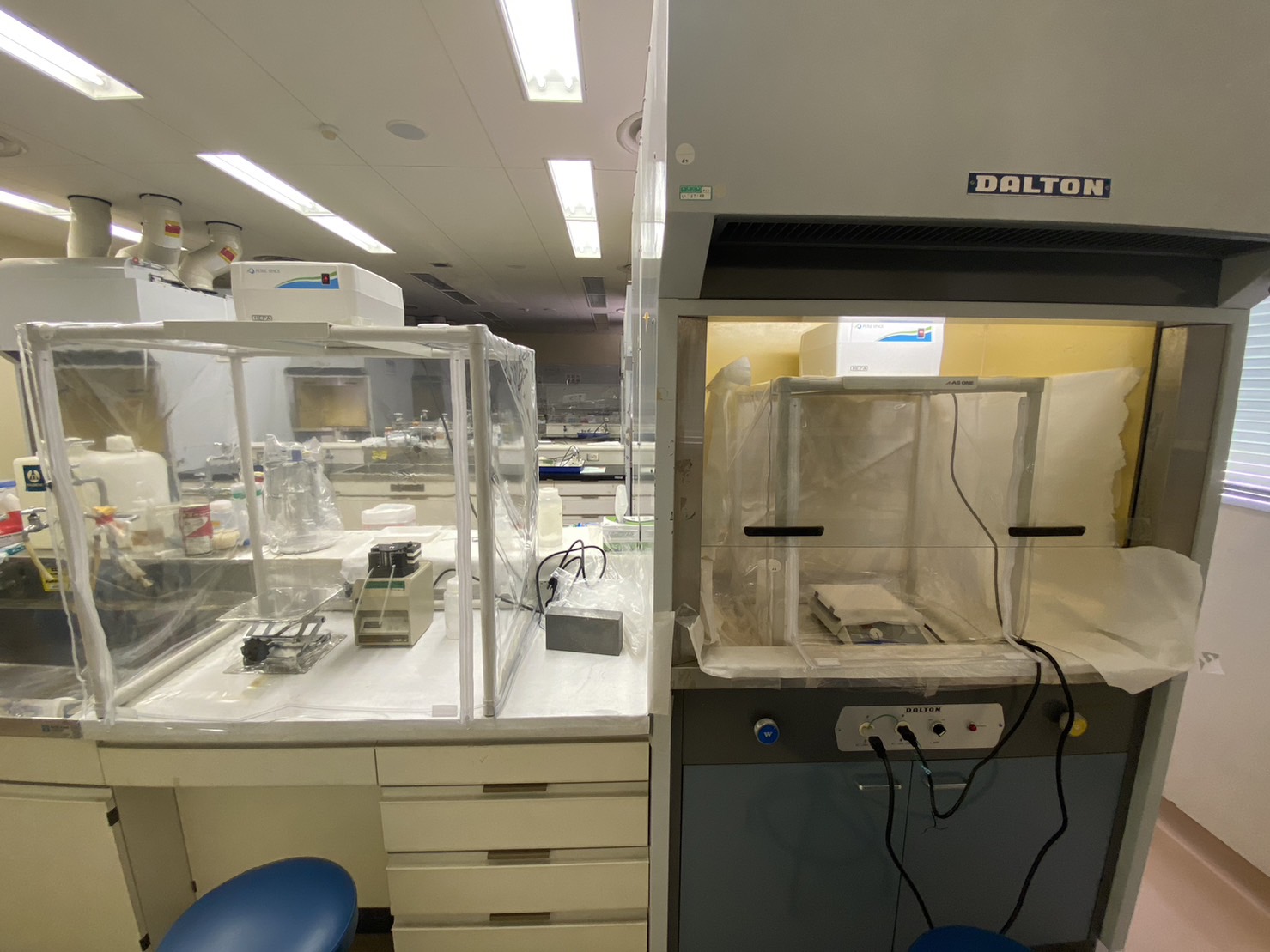}
\caption {A picture of the experimental setup showing the two clean booths used for chemical separation.}
\label{fig:booth}
\end{figure}

The entire procedure is performed in a clean environment to prevent radioactive contamination. Two clean booths with HEPA filters were constructed for this purpose, and the ICP-MS was installed in a separate clean room. Figure~\ref{fig:booth} provides a picture of the location used for the chemical separation. The clean booth on the left is used for the chemical separation of $\Ra$~(see \ref{sec:analig}) and Gd removal~(see \ref{sec:Ln}). The other clean booth, located in a fume hood, is used for heating eluents to evaporate to dryness or condensation~(see \ref{sec:EDTApyrolyzing}).

\subsection {Setup of the ICP-MS}
In the analysis of $\Ra$, the instrumentation includes a triple quadrupole ICP-MS, specifically the Agilent 8800 from Agilent Technologies. Additionally, a solvent removal module, the Aridus~II\texttrademark\ from Teledyne CETAC, minimizes sample loss at the injector and prevents a drop-in plasma temperature. Furthermore, a 100~$\mu$L/min C-Flow PFA nebulizer from Teledyne CETAC was utilized to deliver the sample to the injector. Finally, an ion lens called the s-lens from Agilent Technologies is implemented to optimize the analysis of low-matrix samples.
The differential exhaust section is evacuated using a dual-stage rotary vane vacuum pump, specifically the DS 402 model 9499330M005 from Agilent Technologies.
To enhance the ion transfer rate at the differential exhaust chamber and improve the vacuum, a single-stage rotary vane pump MS40+ model 9499225M008 (Agilent Technologies) was added in series during the measurement process. Following the optimization of several parameters, such as gas flow rates, sampling depth, torch position, and lens bias, the detection limit for $\Ra$ is determined to be 0.06 fg/g (2.2~$\mu$Bq/g, 99.73\% CL.). The demonstrated sensitivity and accuracy of the employed analysis method highlight its effectiveness.

\subsection {$\Ba$ and a HPGe detector}
Barium (Ba) is a commonly used yield tracer to evaluate the collection or recovery rate of Ra~\cite{IAEA}. It has been previously used as a yield tracer in the chemical separation of $\Ra$ by AnaLig$^\text{\textregistered} $ resin\cite{ItoDisk}.
In this study, $\Ba$ ($\tau_{1/2} \sim 10.52$~years) is selected as the yield tracer for $\Ra$.
To measure the 356-keV gamma-rays emitted by radioactive $\Ba$, a high-purity germanium (HPGe) detector (GEM40P4-76, ORTEC) is employed. Using $\Ba$ as the yield tracer and the HPGe detector allows for accurate determination of the recovery rate of $\Ra$.

\section {Method and its performance}

\subsection {Experimental Method} \label{sec:method}
\begin{figure}[!t]
\centering\includegraphics[width=0.9\linewidth] {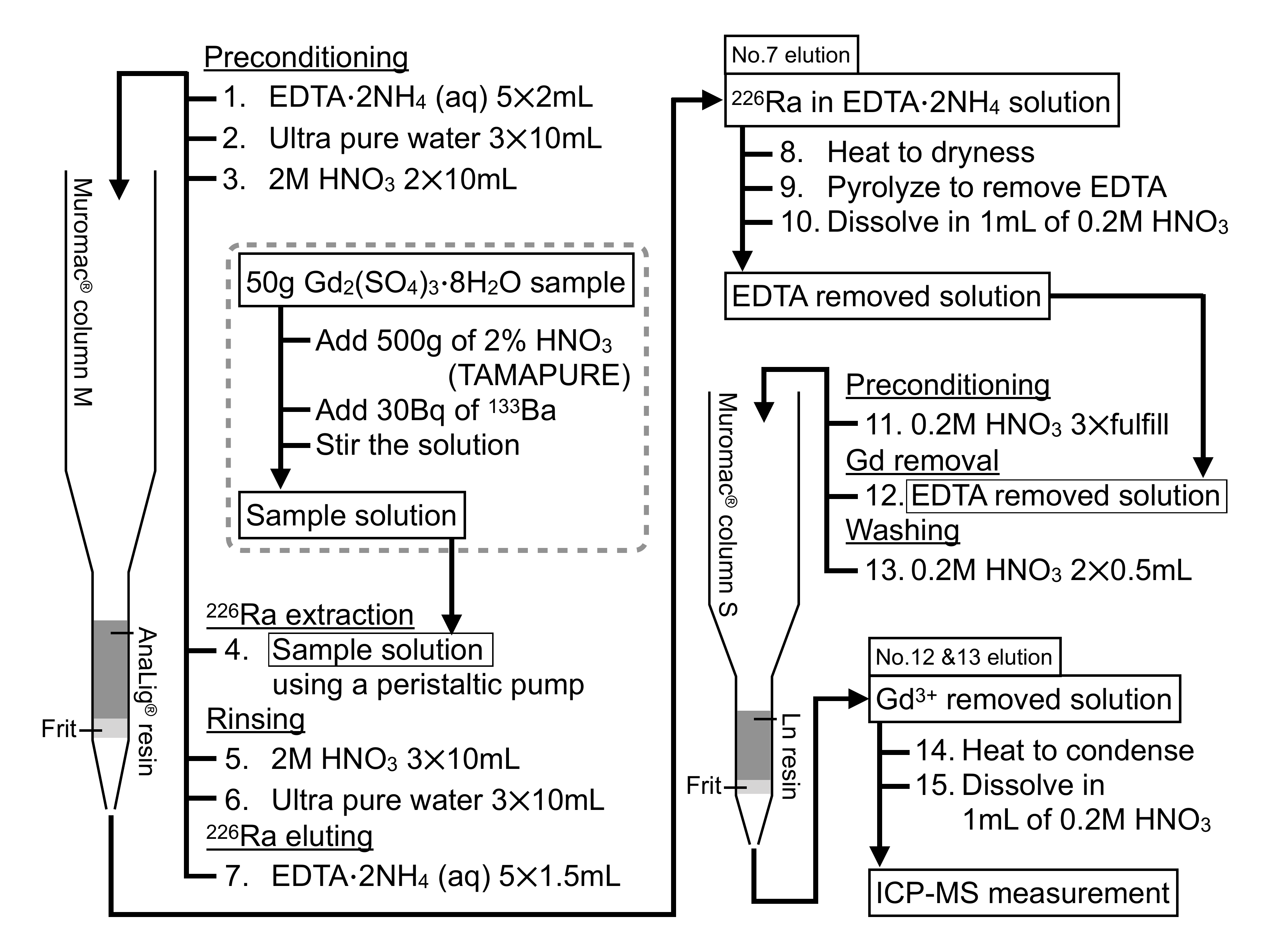}
\caption {The chemical separation process is illustrated in a schematic diagram.}
\label{fig:procedure}
\end{figure}

Figure \ref{fig:procedure} illustrates the chemical separation process.
Before use, all tools are pre-washed with either 10~wt.\% electronic-grade HNO$_3$, 10~wt.\% electronic-grade HCl, 10~wt.\% ultrapure-grade HF or ultrapure water.

\subsubsection {The sample solution preparation} \label{sec:samplemaking}
Following the previous experiments~\cite{SakakiedaProc}, 50~g of $\gdso$ is dissolved in 500~mL of 2 mol/L ultrapure HNO$_3$.
Next, 30~Bq of $\Ba$ is added as a yield tracer for $\Ra$ chemical separation.
And finally, the solution is thoroughly stirred and homogenized.

\subsubsection {Chemical separation of $\Ra$}\label{sec:analig}
To prepare the AnaLig$^\text{\textregistered}$ Ra-01 resin, an approximately 0.03 mol/L EDTA solution is adjusted to be pH 11 and used to pre-wash the resin.
The resin (0.75 mL) is packed on an M-size column.
Then, to condition the resin, the column is flushed with 10 mL of the EDTA solution, followed by 30 mL of ultrapure water and 20 mL of 2 mol/L HNO$_3$.

\begin{figure}[!t]
\centering\includegraphics[width=0.8\linewidth] {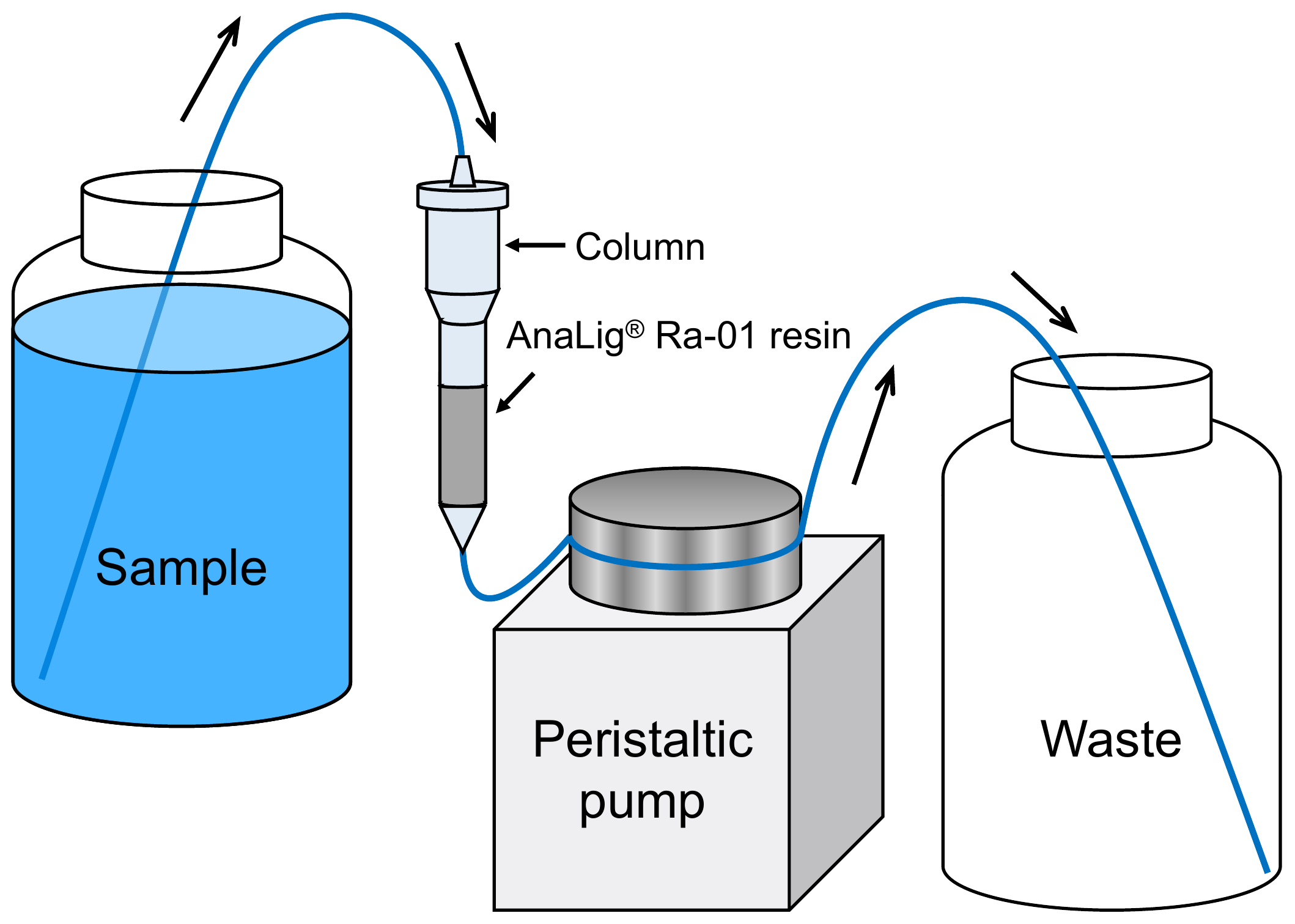}
\caption{A schematic representation of the extraction of $\Ra$ using AnaLig$^\text{\textregistered}$ Ra-01 resin and a peristaltic pump.}
\label{fig:analigsche}
\end{figure}

Fig.~\ref{fig:analigsche} shows a schematic view of the $\Ra$ extraction process using a peristaltic pump. To extract $\Ra$, a peristaltic pump is used to draw the $\gdso$ solution at a flow rate of 5 mL/min through a column filled with AnaLig$^\text{\textregistered}$ Ra-01 resin.

After the $\Ra$ extraction, 30 mL of 2 mol/L ultrapure HNO$_3$ solution is passed through the column to wash out any remaining $\mathrm{Gd^{3+}}$ and $\mathrm{SO_4^{2-}}$ ions.
Next, 30 mL of ultrapure water is passed through the column to replace the solution. Finally, the column passes 7.5 mL of EDTA-2NH$_4$ solution (approximately 0.03 mol/L) at pH 11 to elute $\Ra$, and a quartz beaker collects the eluent.

\subsubsection {EDTA pyrolysis} \label{sec:EDTApyrolyzing}
A large amount of EDTA in the eluent could decrease the sensitivity of the ICP-MS measurement due to the matrix effect.
Therefore, it is necessary to decompose EDTA before measurement.
For the decomposition of EDTA, the eluent obtained after the chemical separation with AnaLig$^\text{\textregistered}$ Ra-01 resin is evaporated to dryness and calcinate it at 600 $^{\circ}$C for three hours using an electric furnace (ROP-001H, AS ONE Corporation.).
After pyrolysis, the residue is dissolved in 1 mL of 0.2 mol/L of ultrapure HNO$_3$.

\subsubsection {Gd removal} \label{sec:Ln}
To ensure accurate $\mathrm{Ra}$ measurements, we need to remove $\mathrm{Gd^{3+}}$ ions from the sample solution, as a concentration of $\mathrm{Gd^{3+}}$ ions greater than $\sim50\mathrm{ppb}$ could decrease $\mathrm{Ra}$ sensitivity in ICP-MS measurements due to matrix effects.
Therefore, Ln resin is used to remove $\mathrm{Gd^{3+}}$ ions. Fig. \ref{fig:procedure} presents a schematic view of the Ln resin column and a scheme for $\mathrm{Gd^{3+}}$ removal using Ln resin. First, an S-size column is packed with approximately 0.25 mL of Ln resin and washed with 6mL of 0.2 mol/L ultrapure HNO$_3$. Next, the eluent undergoing EDTA removal is loaded into the column to remove $\mathrm{Gd^{3+}}$ ions with Ln resin. Next, the column's remaining $\mathrm{Ra}$ is washed with 1mL of 0.2 mol/L ultrapure HNO$_3$. Then, the $\Ra$ eluent is evaporated to dryness and dissolved in 1 mL of 2wt.

\subsubsection {ICP-MS measurement}
Approximately 1mL of the sample eluent is measured using the ICP-MS with the single quad mode selected as the operation mode.
Additionally, the standard addition method is adapted~\cite{SAM}.

\subsection{Performance} \label{sec:blank}
Since the presence of $\Ra$ in the equipment, environment, and procedure can impact the accuracy of the results. Therefore, it is necessary to investigate the effect of these factors.
The $\Ra$ contamination caused by containers, reagents, and environment is referred to as the procedure blank.

Conducting the experimental procedure on a solution of 30 Bq of $\Ba$ in 2 mol/L ultrapure HNO$_3$, without the presence of $\gdso$, allowed the evaluation of the procedure blank.
The resulting elution was measured using ICP-MS to determine the extent of procedure blank contamination.
Table \ref{table:blank} summarizes the measured procedure blanks.
The values presented in the table represent the $\Ra$ concentration in the resulting eluent from the experimental procedure after subtracting the instrumental background.
The instrumental background due to the ICP-MS system can be estimated by measuring a 2wt.\% ultrapure HNO$_3$ solution without solute just before the measurement.
One can convert the ICP-MS measured values to the sample-amount-equivalent concentration of $\Ra$ using the $\Ba$ recovery rate and eluent amount.
We estimated the average procedure blank to be 0.29$\pm$0.01~mBq/kg based on the triple experiments.
This value is used to determine the detection limit, which is found to be 0.32~mBq/kg(99.73\% CL.).

\begin{table}[!t]
  \caption{Summary of procedure blanks measured in Sec.\ref{sec:blank}.
           The average converted concentration of $\Ra$ and the upper limit are 0.29$\pm$0.01 and 0.32~mBq/kg(99.73\% CL.), respectively.}
  \label{table:blank}
  \centering
  \begin{tabular}{cccc}
    \hline
                      & ICP-MS          & $\Ba$         & Converted $\Ra$ \\
                      & measured        & recovery rate & concentration   \\
                      & (fg/g)        & (\%)          & (mBq/kg)        \\
    \hline \hline
    Procedure blank 1 & 0.24$\pm$0.01 & 48.2$\pm$2.4  & 0.37$\pm$0.02   \\
    Procedure blank 2 & 0.13$\pm$0.01 & 46.5$\pm$2.5  & 0.21$\pm$0.01   \\
    Procedure blank 3 & 0.19$\pm$0.01 & 53.2$\pm$1.8  & 0.28$\pm$0.01   \\
    \hline
  \end{tabular}
\end{table}

\section{Application to the SK-Gd samples} \label{sec:result}
SK-Gd loaded approximately 13 tons of $\gdso$ into the ultrapure water in 2020 \cite{T1LoadingPaper}, followed by additional 26 tons in 2022.
The developed method is applied to the last two samples of $\gdso$ from the later loading phase. The two measured samples, listed as lot No.220691(Sample A) and No.220603(Sample B), were the final two lots for the later loading phase and were delivered only one week before loading. Upon arrival, the samples are packed into airtight bags and took approximately 100 g of each sample at the packing stage for analysis.
Using HPGe-detectors, the radioactive impurities other than $\Ra$ are measured in the underground laboratory of the Kamioka Observatory.

The method described in Sec.~\ref{sec:method} was used to analyze two samples, and the resulting $\Ra$ concentration was determined using ICP-MS. After subtracting the instrumental background, Table~\ref{table:SKresult} summarizes the measurements. The recovery rate of $\Ba$ was not measured for the procedure blank in this measurement.
Therefore, the recovery rates measured in the procedure of the two samples were used to convert the ICP-MS measured values into $\Ra$ concentrations.
The left and right values in the converted $\Ra$ concentration and the upper limit of the procedure blank were calculated using the recovery rate of samples A and B, respectively.
As the converted $\Ra$ concentration and the upper limit for the two samples include the procedure blank, the intrinsic $\Ra$ concentration would be lower than these values. SK-Gd has determined that the $\Ra$ contamination levels in the two samples are acceptable, meeting the SK-Gd requirement of $<$0.5~mBq/kg ($\Ra$).

There is a 1.5-fold difference between the procedure blanks measured in this application and those reported in Sec.~\ref{sec:blank}.
Therefore, when using this method to analyze other samples, conducting at least one (preferably three) procedure blank measurements and measuring the sample itself is recommended.

\begin{table}[!t]
  \caption {Results of measuring SK-Gd $\gdso$ samples.
  The converted $\Ra$ concentration and the upper limit of the procedure blank are calculated using the recovery rate of samples A and B, respectively.}
  \label{table:SKresult}
  \centering
  \begin{tabular}{ccccc}
    \hline
                    & ICP-MS          & $\Ba$         & Converted $\Ra$              & upper \\
       Sample       & measured        & recovery rate & concentration                & limit \\
                    & (fg/g)        & (\%)          & (mBq/kg)                     & (99.73\% CL.) \\
    \hline \hline
    Procedure blank & 0.37$\pm$0.02   & -             & 0.89$\pm$0.06, 0.55$\pm$0.03 & 1.06, 0.63 \\
    Sample A        & 0.30$\pm$0.01   & 30.3$\pm$1.28 & 0.84$\pm$0.05                & 0.98 \\
    Sample B        & 0.13$\pm$0.01   & 48.8$\pm$0.76 & 0.24$\pm$0.02                & 0.30 \\
    \hline
  \end{tabular}
\end{table}

\section {Conclusion}
We have developed a new method for rapidly measuring $\Ra$ concentration in $\gdso$ reagent. Using two resins for the chemical separation of $\Ra$ and improving the sensitivity of ICP-MS through a solvent removal module, we can now measure $\Ra$ quickly and with high sensitivity.
Compared to the conventional measurement using an HPGe detector, which takes approximately 20 days, the new method takes only three days to process a batch of samples, including the procedure blank measurement.
The estimated procedure blank was 0.29$\pm$0.01~mBq/kg, and the detection limit, considering both chemical separation and measurement, was 0.32~mBq/kg(99.73\% CL.).

We applied the established method to two $\gdso$ samples from SK-Gd. We found that the amount of $\Ra$ in the samples is within acceptable limits for continuing $^8$B solar neutrino measurements in SK-Gd.
This study can be used where a rapid evaluation of $\Ra$ in $\gdso$ is required.

\section*{Acknowledgement}
This research is supported by the Japan Society for the Promotion of Science (JSPS) KAKENHI Grant-in-Aid for Scientific Research on Innovative Areas, with grant numbers 19H05807 and 20H05243.

\end{document}